\documentclass[a4paper, aps, tightenlines, nofootinbib, 12pt]{revtex4}
\usepackage{hyperref}
\usepackage{amsmath}
 \usepackage{bbm}
\newsavebox{\myhbar}
\savebox{\myhbar}{$\hbar$}
\usepackage{graphicx}
\usepackage{enumerate}
\usepackage{mdframed}
\usepackage{braket}
\usepackage{url}
\usepackage{wasysym}
\usepackage{subfig}
\usepackage{graphicx}
\graphicspath{ {./images/} }
\usepackage{float}
\usepackage{undertilde}
\usepackage{slashed}
\renewcommand*{\hbar}{\mathalpha{\usebox{\myhbar}}}
\begin{document}
%Work and Efficiency of Quantum Moving Heat Engines or 
 \title{The Quantum Otto Heat Engine with a relativistically moving thermal bath} \author {  Nikolaos Papadatos}
 \email{n.papadatos@upatras.gr}
 \affiliation{ Department of Physics, University of Patras, 26500 Greece}

\begin{abstract}
We investigate the quantum thermodynamic cycle of a quantum heat engine carrying out an Otto thermodynamic cycle. We use the thermal properties of a moving heat bath with relativistic velocity with respect to the cold bath. As a working medium, we use a two-level system and a harmonic oscillator that interact with a moving heat bath and a static cold bath. In the current work, the quantum heat engine is studied in the high and low temperatures regime. Using quantum thermodynamics and the theory of open quantum system we obtain the total produced work, the efficiency and the efficiency at maximum power.  The maximum efficiency of the Otto quantum heat engine depends only on the ratio of the minimum and maximum energy gaps. On the contrary, the efficiency at maximum power and the extracted work decreases with the velocity since the motion of the heat bath has an energy cost for the quantum heat engine. Furthermore, the efficiency at maximum power depends on the nature of the working medium.
\end{abstract}

\maketitle

\section{Introduction}
\subsection{Motivation}

The aim of this paper is to study the quantum heat engines (QHEs) \cite{Alicki} in which one bath moves with relativistic velocity $u$ with respect to the other. In particular, we study the quantum equivalent of the classical Otto cycle which has been studied extensively in recent years \cite{Abah0}-\cite{Leggio}. We show that the Otto QHE is less efficient because of the relativistic motion of the heat reservoir.

This work is connected with the long standing debate about the correct relativistic transformation of temperature. This debate has its origin to the fact that a moving observer in a thermal reservoir can not detect Black-body radiation. Up to now, several approaches have been made but there is no consensus to the question if the temperature of a body in relativistic motion is lower, greater or equal with respect to the same body at rest---for reviews see, Ref. \cite{reviews}

The first researchers who analyzed Lorentz transformations for a body of temperature $T$ in its rest frame was Von Mosengeil \cite{vMos},  Planck \cite{Planck} and Einstein \cite{Einstein}. They proposed that a moving observer with constant velocity $u$ detects a temperature $T' = T\sqrt{1-u^2} < T$. On the contrary, Ott supported that $T' = T/\sqrt{1-u^2} > T$ \cite{Ott}. Arzeliès agreed with Ott's formula, but suggested a different law of transformation for the internal energy \cite{Arzel}. Later, Landsberg stressed that the temperature is a Lorentz scalar, i.e.,  $T' = T$ \cite{Landsberg}, \cite{CaSa}.

In this work, we approach the issue of relativistic transformation of temperature  in the context of QHEs, in which one of the two baths moves with relativistic velocity $u$ with respect to the other. The QHE was first studied by Scovil and Schulz-DuBois in 1959 using a 3-level maser \cite{Scovil}. Generally speaking, QHE produces work using as a working medium a quantum system. Every QHE has a characteristic quantum thermodynamic cycle that consists of a quantum adiabatic process or a quantum isochoric process and a quantum isothermal process. In the adiabatic process, the working medium is isolated from the thermal reservoir. On the contrary, in the quantum isochoric and isothermal process, the working medium is weakly coupled to thermal reservoir. 
 
 In recent years, QHEs have been studied in a variety of thermal heat baths. In particular, in Refs. \cite{Klaers}-\cite{Assis} a squeezed thermal reservoir was used. Moreover, in Ref. \cite{Abah1} the QHE operates between nonequilibrium stationary reservoirs and in Ref. \cite{Alickiengine} the QHE consists of a two-level system that is permanently, weakly, coupled to two separated heat baths at different temperatures. Furthermore, in Ref. \cite{Scullycoherent} the QHE is coupled to quantum coherent reservoir and in Ref. \cite{Dillenschneidercorrelation} to quantum correlated reservoir. For the first time the creation of a QHE by using the properties of a moving heat bath with relativistic velocity $u$ is presented in this paper.
 
 Our work \cite{nikos} provides a link between relativistic motion of the thermal heat bath and the QHEs. In that reference, we analysed the thermodynamics of a quantum system that moves with constant velocity $u$ with respect to the heat bath. The master equation for the reduced dynamics of the moving quantum system was extracted using a coupling of the Unruh-DeWitt type and a coupling that involves the derivative of the field with respect to time. Moreover, it had the form of the quantum-optical master equation  \cite{BrePe07},  \cite{Eitan Geva and Ronnie Kosloff}, \cite{Yair Rezek and Ronnie Kosloff} but the mean number of quanta did not have the usual Planckian form. Finally, the second law of thermodynamics led to a surprising equivalence: a moving heat bath is physically equivalent to a mixture of heat baths at rest, each with a different temperature.

The master equation connects the relativistic motion of the thermal heat bath and the QHE. Specifically, the first term of the master equation describes the unitary evolution of the density operator, namely the quantum adiabatic process and the second term is the Lindbladian which describes the non-unitary evolution of the density operator, i.e., the thermalization process.

As far as the ideal QHEs are concerned, they are characterized from a reversible process, where the increase of entropy is nearly zero. On the contrary, zero entropy production is not a necessary condition for reversibility \cite{entropyproductionLee}. In particular, reversible process is an infinitely slow process evolving close to thermodynamic equilibrium. Moreover, we have complete conversion from one form of work to another, for example heat can be completely converted  into work and the outpower power is zero \cite{Callen}. Consequently, from a practical point of view, an important issue is to optimize the performance of QHEs and to achieve maximum output power. We show that the efficiency at maximum power decreases with the velocity of the thermal heat bath.

Among the first who study the efficiency at maximum power were Curzon and Ahlborn (CA) \cite{Curzon}. Using the endo-reversible approximation, i.e, neglecting the dissipation in the auxiliary system, they optimized the Carnot cycle with respect to power and they found that the efficiency at maximum power is $\eta_{CA}=1-\sqrt{\frac{T_c}{T_h}}=1-\sqrt{1-\eta_C}=\frac{\eta_C}{2}+\frac{\eta^2_C}{8}+..$ where $\eta_{C}$ the Carnot efficiency. At first order in $\eta_{C}$,  the efficiency at maximum power is half of the Carnot efficiency \cite{Broeck}. Here, we show that the efficiency at maximum power of CA is reproduced for $u\rightarrow 0$ when it comes to the harmonic oscillator as the working medium.

\subsection{Analysis and results}
Our QHE consists of a small quantum system as the working medium that is weakly coupled to a hot and a cold bath on alternate pace according to the Otto thermodynamic cycle.  The hot bath moves with relativistic velocity $u$ with respect to the cold bath. As a working medium we use a two-level system and a harmonic oscillator that undergoes four strokes. From the four strokes, the two strokes are responsible for procedure of thermalization. With the use of the Born-Markov approximations the thermalization quantum master equation is produced \ref{masterequation}. The rest two strokes are adiabatic transitions where $\hat{\rho}(t)$ remains at the initial equilibrium state due to the fact that $[\hat{H},\hat{\rho}_{\beta,\omega}]=0$ for all t.

Our results are the following:
\begin{enumerate}[(i)]
	\item The efficiency of the QHE depends only on the ratio of the minimum and maximum energy gaps. Moreover, using a classical thermal bath the same efficiency is found \cite{Kieu}, which implies a universal bound that does not depend on the motion of the thermal reservoir.

	\item  Maximum work depends on the temperature of the two baths and it decreases when $T_c\rightarrow T_h$, see Fig. 1 and Fig. 3.
	
		\item  The performance of QHE, i.e., the efficiency at maximum power and the total produced work depends on the nature of the working medium, see Fig. 1-4.
		
	\item The efficiency at maximum power and the total produced work, in both the high and low temperature regime, decreases with the velocity, see Fig. 1-4. This means that the motion of the heat bath has an energy cost for the QHE. In the limit $u\rightarrow 0$ the results of Ref. \cite{ Abah} are reproduced.

	\item The temperature determines the classical or quantum behaviour of the QHE. In particular, in the low temperature regime, the system has quantum nature because the thermodynamic quantities depend on $\hbar$ only through the combination $\hbar\omega$.
	
    \item It is impossible to define an effective temperature so that the efficiency at maximum power of QHE  that uses a moving heat bath with relativistic velocity $u$, is equivalent to the efficiency at maximum power of QHE that uses a heat bath at rest.

\end{enumerate}

The structure of this paper is the following. In Sec.\ref{Description of the system} we describe the necessary thermodynamic quantities in order to study the QHE. In Sec. \ref{Quantum Otto heat engine}, we examine the quantum Otto heat engine with working medium a two-level system and a harmonic oscillator. In Sec. \ref{Effective Temperature} we try to define an effective temperature. The conclusions are presented in Sec. \ref{Conclusions}.

\section{Description of the system}
\label{Description of the system}

The system under investigation is composed of a microscopic probe S and a quantum scalar field. The combined system is described by a Hilbert space ${\cal H}_S \otimes {\cal H}_{\phi}$, where ${\cal H}_S $ is the Hilbert space of the probe and ${\cal H}_{\phi}$ is the same as before of the field.

The Hamiltonian is

\begin{eqnarray}
	\hat{H}=\hat{H}_{S}\otimes\hat{I}_{\phi}+\hat{I}_S\otimes\hat{H}_{\phi}+\hat{H}_{int} \label{Ham1}
\end{eqnarray}

where $\hat{H_S}$ is the Hamiltonian of S, $\hat{H}_{\phi}$ is the Hamiltonian of a free massless scalar field 
\begin{equation}
	\hat{H}_\phi= \frac{1}{2} \int d^3x\Big(\hat{\pi}^2+(\nabla\hat{\phi})^2\Big)
\end{equation}
where $\hat{\pi}({\pmb x})$ is the conjugate momentum of the field $\hat{\phi}({\pmb x})$. $\hat{H}_{int}$ is the Hamiltonian for the system-free massless scalar field interaction; finally $\hat{I}_{\phi}$ and $\hat{I}_{S}$ are the identities in the corresponding Hilbert spaces.\\
 The operators in the interaction picture are defined as follows:
 \begin{align}
\label{interaction hamiltonian}
\hat{H}_I=e^{\frac{i}{\hbar}(\hat{H}_S+\hat{H}_{\phi})\tau}\hat{H}_{int}e^{-\frac{i}{\hbar}(\hat{H}_S+\hat{H}_{\phi})\tau}
\end{align}
\begin{align}
\hat{\rho}_{I}=e^{\frac{i}{\hbar}(\hat{H}_S+\hat{H}_{\phi})\tau}\hat{\rho}_{tot}e^{-\frac{i}{\hbar}(\hat{H}_S+\hat{H}_{\phi})\tau}
\end{align}

The Born-Markov master equation is,
\cite{BrePe07}
\begin{align}
	\label{time evolution density matrix}
	\frac{d\hat{\rho}_{I,S}(\tau)}{d\tau}=-\frac{1}{\hbar^2}Tr_{B}\int_{0}^{\infty}\Big[\hat{H}_I(\tau),[\hat{H}_I(\tau-s),\hat{\rho}_S(\tau)\otimes \hat{\rho}_B]\Big]ds
\end{align}

{\em Green function} for a quantum field $\hat{\phi}(x)$ is the mean value of $\hat{\phi}(\tau,\pmb x)$, $\hat{\phi}(s,\pmb x')$. 
The {\em thermal Wightman function} are the thermal correlators \cite{Wel00}:

\begin{eqnarray}
	G^+_{\beta}(\tau,\pmb x;s,\pmb x') = Tr \left[ \hat{\phi}(\tau,\pmb x) \hat{\phi}(s,\pmb x')\hat{\rho}_{\phi} \right], \label{Wightman+}\\
		G^-_{\beta}(\tau,\pmb x;s,\pmb x') = Tr \left[   \hat{\phi}(s,\pmb x')\hat{\phi}(\tau,\pmb x)\hat{\rho}_{\phi} \right]. \label{Wightman-}
\end{eqnarray}

We assume an initial state $\hat{\rho}_0 \otimes \hat{\rho}_{\phi}$ where
\begin{eqnarray}
	\hat{\rho}_{\phi} = \frac{e^{- \beta \hat{H}_{\phi}}}{Tr (e^{- \beta \hat{H}_{\phi}})}
\end{eqnarray}
is a Gibbs state of temperature $\beta^{-1}$. Due to the fact that the spacetime is static and the system is invariant under spatial translations and rotations, these functions depend only on  $r = |{\pmb x}-\pmb x'|$  \cite{Letaw}.

\begin{eqnarray}
	G^+_{\beta}(\tau,\pmb x;s,\pmb x') = G^+_{\beta}(\tau-s,r)\\
	G^-_{\beta}(\tau,\pmb x;s,\pmb x') = 	G^-_{\beta}(\tau-s,r)
\end{eqnarray}

We find that the thermal Wightman function takes the form

 \begin{eqnarray}
 G(x) = G_0(x) + \frac{1}{4 \pi^2r} \int_0^{\infty} dk n_{k}\left[\sin[k(t+r)] - \sin[k(t-r)]\right], \label{wightman}
 \end{eqnarray}
The first term is the Wightman function of vacuum
\begin{eqnarray}
	G_0(x) = - \lim_{\epsilon \rightarrow 0^+} \frac{1}{4\pi^2\left[ (t- i \epsilon)^2-r^2\right]},
\end{eqnarray}  and $n_k$ is the expected number of particles of momentum ${\pmb k}$. Furthemore, since the system is invariant under spatial translations and rotations, $n_k$ depends only on  $k = |{\pmb k}|$ due to the spherical symmetry of the system. For a Gibbsian field state, $n_k = (e^{\beta k}-1)^{-1}$.

Studying the following trajectories of the detector \cite{Unruh76, Dewitt, HLL12}
\begin{equation}
	\label{trajectories}
	x(\tau)=\Big(\frac{1}{\sqrt{1-u^2}},\frac{u}{\sqrt{1-u^2}},0,0\Big)\tau
\end{equation}

the correlation function becomes

\begin{eqnarray}
	g_{UdW}(\tau) = - \lim_{\epsilon \rightarrow 0^+} \frac{1}{4\pi^2 (\tau- i \epsilon)^2 } + \frac{\sqrt{1-u^2}}{4 \pi^2 |\tau|  u  } \int_0^{\infty} dk n_{k}\left[\sin\Big(\sqrt{\frac{1+u}{1-u}}k \tau\Big) - \sin\Big(\sqrt{\frac{1-u}{1+u}}k \tau\Big)\right]. \label{gt1}
\end{eqnarray}

 The second-order master equation of the reduced density matrix $\hat{\rho}$ of the probe in the original picture,
\begin{eqnarray}
\frac{\partial \hat{\rho}}{\partial \tau} = - \frac{i}{\hbar} [ \hat{H}, \hat{\rho}] +   \lambda^2 \sum_{\omega} \tilde{g}(\omega)  \left( \hat{A}_{\omega}\hat{\rho}\hat{A}^{\dagger}_{\omega}
- \hat{A}^{\dagger}_{\omega} \hat{A}_{\omega}\hat{\rho} \right) \nonumber +\\
+\lambda^2 \sum_{\omega} \tilde{g}^*(\omega) \left(  \hat{A}^{\dagger}_{\omega}\hat{\rho}\hat{A}_{\omega}  - \hat{\rho} \hat{A}^{\dagger}_{\omega} \hat{A}_{\omega}\right),
\end{eqnarray}

where,
\begin{eqnarray}
\tilde{g}(\omega) = \int_0^{\infty}  d \tau e^{i \omega \tau} g(\tau).
\end{eqnarray}

We obtain the master equation
\begin{eqnarray}
		\label{masterequation}
	\frac{\partial \hat{\rho}}{\partial \tau} = - \frac{i}{\hbar} [ \hat{H} + \hat{H}_{LS}, \hat{\rho}] +    \sum_{\omega > 0 } \gamma(\omega) [N(\omega)+1]  \left[ \hat{A}_{\omega}\hat{\rho}\hat{A}^{\dagger}_{\omega} -\frac{1}{2}  \hat{A}^{\dagger}_{\omega}\hat{A}_{\omega}\hat{\rho}  -  \frac{1}{2}  \hat{\rho} \hat{A}^{\dagger}_{\omega} \hat{A}_{\omega}\right]\nonumber \\
	+     \sum_{\omega > 0 } \gamma(\omega) N(\omega)  \left[ \hat{A}^{\dagger}_{\omega}\hat{\rho}\hat{A}_{\omega} -\frac{1}{2}  \hat{A}_{\omega}\hat{A}^{\dagger}_{\omega}\hat{\rho}  -  \frac{1}{2}  \hat{\rho} \hat{A}_{\omega} \hat{A}^{\dagger}_{\omega}\right] \label{mastereq}
\end{eqnarray}
where $	N(\omega)$ is the number of the quanta
\begin{eqnarray}
	\label{N}
	N(\omega,\beta,u) = \frac{\sqrt{1-u^2}}{2u\beta \hbar\omega} \log \Bigg(\frac{ 1 - e^{-\beta \hbar\omega \sqrt{\frac{1+u}{1-u}}}}{1 - e^{-\beta \hbar\omega \sqrt{\frac{1-u}{1+u}}}}\Bigg)
\end{eqnarray}

Due to the fact that the particle number distribution is not of the Planckian form of Black-body radiation Landsberg and Matsas \cite{CoMa}-\cite{LaMa1} concluded that it is impossible to define a universal relativistic transformation of temperature ---see, also \cite{Naka} for a critique. 

In Eq.(\ref{N}) if we take the limit $u\rightarrow 0$ the master equation coincides with the quantum optical master equation \cite{BrePe07}, namely
\begin{eqnarray}
	 n(\omega,\beta)= \frac{ 1 }{ e^{\beta\hbar \omega}-1}. 
\end{eqnarray}

\subsection{Two-level atom}
The Hamiltonian of a two-level atom of frequency $\Omega_0$ is $\hat{H} = \frac{1}{2} \hbar\Omega_0 \hat{\sigma}_z$ and the coupling operator $\hat{A} = \hat{\sigma}_1$. Replacing the transition operators: $\hat{A}_{\Omega_0}\rightarrow\hat{\sigma}_-$ and $\hat{A}_{-\Omega_0}\rightarrow\hat{\sigma}_+$ we obtain the master equation
\begin{eqnarray}
	\frac{\partial \hat{\rho}}{\partial \tau} = - i \Omega [ \hat{\sigma}_3, \hat{\rho}] + \Gamma_0  [N(\Omega_0)+1]  \left( \hat{\sigma}_-\hat{\rho}\hat{\sigma}_+  -\frac{1}{2}  \hat{\sigma}_+\hat{\sigma}_- \hat{\rho}  -  \frac{1}{2}  \hat{\rho}  \hat{\sigma}_+\hat{\sigma}_-  \right)\nonumber \\
	+   \Gamma_0 N(\Omega_0)  \left(   \hat{\sigma}_+ \hat{\rho}\hat{\sigma}_-  -\frac{1}{2}  \hat{\sigma}_- \hat{\sigma}_+ \hat{\rho}  -  \frac{1}{2}  \hat{\rho} \hat{\sigma}_- \hat{\sigma}_+\right),
\end{eqnarray}
where $\Gamma_0 := \gamma(\Omega_0)$ is the decay coefficient for the atom in vacuum and $\Omega = \Omega_0 + 2 \Delta(\Omega_0)$ is the Lamb-shifted excitation frequency.

There is a unique asymptotic state
\begin{eqnarray}
		\label{asymptoticTLS}
	\rho(\tau) = \frac{1}{2 N(\Omega_0)+1 }
	\bordermatrix{&              &              \cr
		& N(\Omega_0)
		& 0 \cr
		& 0
		& N(\Omega_0) + 1\cr}.
\end{eqnarray}
The expectation value of  energy is

\begin{eqnarray}
	\label{energyTLS}
	\langle \hat{h} \rangle = - \frac{ \hbar\Omega_0}{2 [2N(\Omega_0)+1]}.
\end{eqnarray}

\subsection{Harmonic oscillator}
The Hamiltonian of a harmonic oscillator of mass $m$ and frequency $\omega_0$ is $\hat{H} = \hbar\omega_0 (\hat{a}^{\dagger}\hat{a}+\frac{1}{2})$ and  $\hat{x} = \sqrt{\frac{\hbar}{2m\omega_0}}(\hat{a}+\hat{a}^{\dagger})$, $\hat{p} = -i\sqrt{\frac{m\hbar\omega_0}{2}}(\hat{a}-\hat{a}^{\dagger})$ are the operators associated with position and momentum. Replacing the transition operators $\hat{A}_{\omega_0} \rightarrow \sqrt{\frac{\hbar}{2m\omega_0}} \hat{a}$ and $\hat{A}_{-\omega_0} \rightarrow   \sqrt{\frac{\hbar}{2m\omega_0}}\hat{a}^{\dagger}$ the master equation becomes

\begin{align*}
	\frac{d\hat{\rho}(\tau)}{d\tau}
	&=-i\omega_0[\hat{a}^\dagger\hat{a},\hat{\rho}]+\Gamma_0\Big( N(\omega_0) + 1 \Big)\Big(\hat{a}\hat{\rho}\hat{a}^\dagger-\frac{1}{2} \hat{a}^\dagger\hat{a}\hat{\rho}-\frac{1}{2}\hat{\rho}\hat{a}^\dagger\hat{a}\Big)+\\
	& +\Gamma_0 N(\omega_0) \Big(\hat{a}^\dagger\hat{\rho}\hat{a}-\frac{1}{2}\hat{a}\hat{a}^\dagger\hat{\rho} - \frac{1}{2}\hat{\rho}\hat{a}\hat{a}^\dagger\Big),
\end{align*}
where $\Gamma_0 = \frac{\gamma (\omega_0)}{2m \omega_0}$.

The asymptotic state with matrix elements in the energy basis is,
\begin{eqnarray}
	\rho_{nn'} =\frac{1}{N(\omega_0)+1}\Bigg(\frac{N(\omega_0)}{N(\omega_0)+1}\Bigg)^n \delta_{nn'},
\end{eqnarray}
with mean energy
\begin{eqnarray}
		\label{energyHO}
	\langle \hat{H} \rangle = \hbar \omega_0\Big(N(\omega_0)+\frac{1}{2}\Big).
\end{eqnarray}

\section{Quantum Otto heat engine}
\label{Quantum Otto heat engine}
\subsection{Two-level system}

The QHE is comprised of a two-level system (qubit) as the working medium. It interacts with a heat bath of relativistic velocity $u$ with respect to the cold bath so as to perform a quantum Otto thermodynamic cycle. \\
The general quantum Otto cycle has the following four repeated cyclically steps:

\begin{enumerate}[(i)]
	
	\item Isentropic compression:   $A(\omega_c,\beta_c)\rightarrow B(\omega_h,\beta_c)$. Initially, the qubit is isolated and its evolution is unitary and quasistatic. The unitary evolution leads the qubit's state from $\rho_A \rightarrow \rho_B=U\rho_A U^{\dagger}$, where $U$ is the unitary operator and $\rho $ is given from \ref{asymptoticTLS}. The  qubit's Hamiltonian from $\hat{H}_A=\frac{1}{2}\hbar \omega_c \hat{\sigma}_z$ to  $\hat{H}_B=\frac{1}{2}\hbar \omega_h \hat{\sigma}_z$.
	\item Hot isochore: $B(\omega_h,\beta_c)\rightarrow C(\omega_h,\beta_h)$. In the second stroke, the qubit interacts with the moving thermal bath at temperature $T_h$. We assume that the qubit after thermalization will be in thermal equilibrium with the heat bath at temperature $T_h$. The Hamiltonian of the qubit $H_B$ does not change.
	
	\item  Isentropic expansion:    $C(\omega_h,\beta_h)\rightarrow D(\omega_c,\beta_h)$. In the third stroke, the qubit is again isolated and its evolution is unitary and quasistatic. The frequency returns to the initially value $\omega_c$ having the following transitions, $\rho_C \rightarrow \rho_D=\tilde{U}\rho_C \tilde{U}^{\dagger} $ and $H_C=\frac{1}{2}\hbar \omega_h \sigma_z$ to  $H_D=\frac{1}{2}\hbar \omega_c \sigma_z$.
	
	\item  Cold isochore:   $ D(\omega_c,\beta_h)\rightarrow A(\omega_c,\beta_c)$. Finally, at constant frequency the qubit interacts with static cold bath at temperature $T_c$. We assume that the qubit after thermalization will be in thermal equilibrium with the cold bath at temperature $T_c$. The Hamiltonian of the qubit $H_D$ does not change.
\end{enumerate}

The interaction of the working medium with the thermal bath is governed by the master equation and the expectation value of energy for every stroke is computed from \ref{energyTLS} and (\ref{N}). Thus,

\begin{align}
	\begin{split}
			\langle \hat{H} \rangle_A &= - \frac{ \hbar\omega_c}{2 [2n(\omega_c,\beta_c)+1]},  \langle \hat{H} \rangle_B =  - \frac{ \hbar\omega_h}{2 [2n(\omega_c,\beta_c)+1]}.\\
				\langle \hat{H} \rangle_C& = - \frac{ \hbar\omega_h}{2 [2N(\omega_h,\beta_h,u)+1]},  \langle \hat{H} \rangle_D = - \frac{\hbar \omega_c}{2 [2N(\omega_h,\beta_h,u)+1]}.
	\end{split}
\end{align}

We compute the expectation value of heat and work for every stroke:

\begin{align}
	\begin{split}
		\langle \hat{W} \rangle_{AB} &=   \frac{ \hbar(\omega_c-\omega_h)}{2 [2n(\omega_c,\beta_c)+1]},\\
		\langle \hat{Q} \rangle_{H} &=    \frac{\hbar\omega_h}{2}\Big(\frac{ 1}{ [2n(\omega_c,\beta_c)+1]} - \frac{ 1}{ [2N(\omega_h,\beta_h,u)+1]}\Big),\\
		\langle \hat{W} \rangle_{CD}& =  \frac{\hbar (\omega_h-\omega_c)}{2 [2N(\omega_h,\beta_h,u)+1]},\\
		\langle \hat{Q} \rangle_{C} &= \frac{\hbar\omega_c}{2}\Big(- \frac{ 1}{ [2n(\omega_c,\beta_c)+1]}+  \frac{ 1}{ [2N(\omega_h,\beta_h,u)+1]}\Big).
	\end{split}
\end{align}

The average heat $\langle \hat{Q} \rangle_{H}\geq 0$ absorbed
from the hot thermal reservoir and the average heat $\langle \hat{Q} \rangle_{C}\leq 0$ absorbed from the cold thermal reservoir. The mean work $\langle \hat{W} \rangle_{ab}\geq 0$ expresses the consumed work by the system and the $\langle \hat{W} \rangle_{cd}\leq 0$ means that the system produces work.

The efficiency of the QHE is the proportion of the total work
per cycle with respect to the heat absorbed from the hot thermal reservoir:

\begin{align}
	\label{etaottoTLS}
	&\eta = -\frac{ \langle \hat{W} \rangle_{total}}{\langle \hat{Q} \rangle_{H}}=1-\frac{\omega_c}{\omega_h}
\end{align}

We notice that the efficiency does not depend on the velocity of the hot bath. Furthermore, using a classical thermal bath the same efficiency is found \cite{Kieu} which implies a universal bound that does not depend on the motion of the thermal reservoir.

\subsubsection{Total work}
From a practical point of view,  when it comes to real QHEs, the power output  $P=- \langle \hat{W} \rangle_{total}/t_{total}$ and the efficiency at maximum power  is a necessity \cite{Andresen}. The power output is the produced work per unit time and is zero for ideal engines such as Carnot, Otto, Diesel, etc\cite{Callen}.\\
We want to find the efficiency at maximum power output for given bath temperatures. The total produced work is:

\begin{align}
\langle W\rangle=\frac{\hbar(\omega_c-\omega_h)}{2}\Big(\frac{1 }{ 2n(\omega_c,\beta_c)+1}-\frac{ 1}{ 2N(\omega_h,\beta_h,u)+1}\Big)
\end{align}

Let us examine the {\em high temperature limit: $\beta \omega \rightarrow 0$}, the total produced work is:
\begin{align}
	\begin{split}
\langle W_{ht}\rangle&=\frac{ \hbar^2(\omega_c-\omega_h)\Big(\sqrt{1-u^2}\log(\frac{1+u}{1-u})\beta_c\omega_c-2u \beta_h\omega_h\Big)}{4\sqrt{1-u^2}\log(\frac{1+u}{1-u})}
	\end{split}
\end{align}

From (\ref{etaottoTLS}) we obtain:\\

\begin{align}
	\begin{split}
		\langle W_{ht}\rangle&=\frac{ \eta\hbar^2\omega_c^2\Big(\sqrt{1-u^2}(\eta-1)\log(\frac{1+u}{1-u})\beta_c+2u \beta_h\Big)}{4\sqrt{1-u^2}(\eta-1)^2\log(\frac{1+u}{1-u})}
	\end{split}
\end{align}

\begin{figure}[H]
	\centering
	{{\includegraphics[width=15cm]{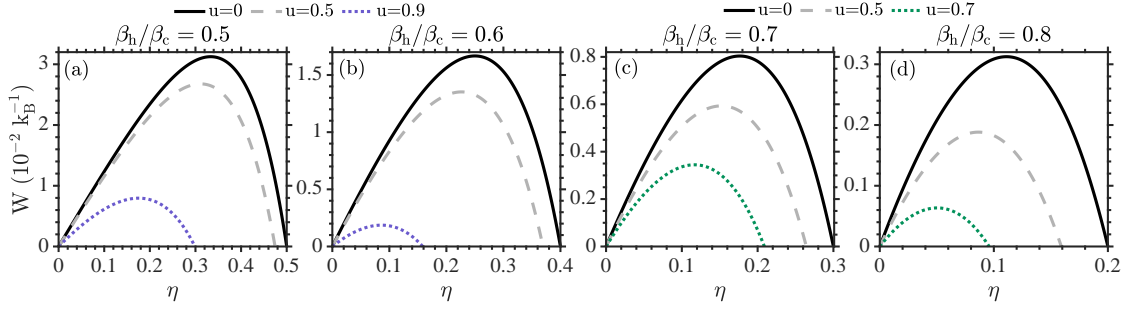} }}%
	% \caption{2 Figures}%
	\caption{\small  Total produced work with respect to efficiency for different ratios of temperatures of reservoirs $\frac{\beta_h}{\beta_c}$ and velocities of the heat bath.}
\end{figure}

In Fig. 1, it is shown that the total produced work decreases with the velocity of the heat bath and when $T_c\rightarrow T_h$. We want to find the maximum value of work with respect to $\omega_h$, thus, we assume that $\beta_c, \omega_c, \beta_h$ are constants and the only variable is $\omega_h$.
\begin{align}
	\begin{split}
		\frac{dW_{ht}}{d\omega_h}&=0.
	\end{split}
\end{align}
Consequently

\begin{align}
	\begin{split}
		\frac{\omega_h}{\omega_c}&=\frac{\Big(\sqrt{1-u^2}\log(\frac{1+u}{1-u})\beta_c+2u \beta_h\Big)}{4u\beta_h}
	\end{split}
\end{align}

The efficiency at maximum work, from (\ref{etaottoTLS}) is

\begin{align}
	\begin{split}
		&\eta^{mw}_{ht} = 1-\frac{4u\beta_h}{\sqrt{1-u^2}\log(\frac{1+u}{1-u})\beta_c+2u \beta_h  }
	\end{split}
\end{align}
and for $u\rightarrow 0$,
\begin{align}
	\begin{split}
		&\eta^{mw}_{ht,u\rightarrow 0} = \frac{\beta_c-\beta_h}{\beta_c+\beta_h}
	\end{split}
\end{align}

\begin{figure}[H]
	\label{fig2}
	\centering
	{{\includegraphics[width=7cm]{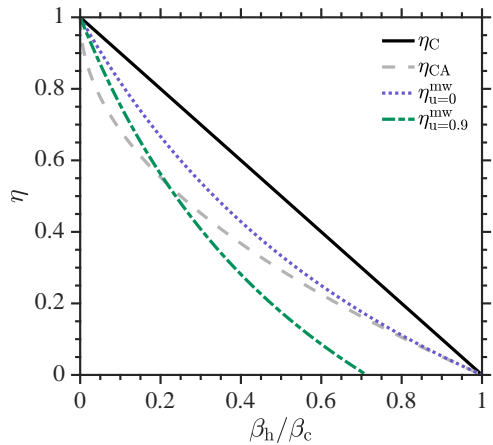} }}%
	% \caption{2 Figures}%
	\caption{ \small  Efficiency at maximum power with respect to $\beta_h/\beta_c$ in comparison with the Carnot limit $\eta_C$ and efficiency of CA $\eta_{CA}$.}
\end{figure}

In Fig. 2, we plot the efficiency at maximum power with respect to the ratio of minimum and maximum temperature. We notice that when the velocity of the heat bath is zero, the efficiency at maximum power surpasses the efficiency of CA. Moreover, the efficiency at maximum power decreases with the velocity $u$ since the motion of the heat bath has an energy cost for the QHE.

Let us now, proceed to examine the {\em  low temperature limit: $\beta_c \omega_c \rightarrow \infty$}, the total produced work is:

\begin{align}
	\langle W_{lt}\rangle=\frac{\hbar(\omega_c-\omega_h)}{2}\Big(1-\frac{ u\beta_h \hbar\omega_h}{ \sqrt{1-u^2}\log(\frac{1+u}{1-u})}\Big)\\
\end{align}

Following the same procedure, as previously mentioned, the efficiency at maximum work in the low temperature regime, from (\ref{etaottoTLS}) is

\begin{align}
	\begin{split}
		&\eta^{mw}_{lt} = 1-\frac{\omega_c}{\omega_h}=\frac{\sqrt{1-u^2}\log(\frac{1+u}{1-u}) -u\beta_h\hbar\omega_c}{\sqrt{1-u^2}\log(\frac{1+u}{1-u}) +u\beta_h\hbar\omega_c}
	\end{split}
\end{align}
for \begin{align}
	\begin{split}
		\omega_h=\frac{1}{2}\Big( \frac{\sqrt{1-u^2}\log(\frac{1+u}{1-u}) }{u\hbar\beta_h}+\omega_c\Big)
	\end{split}
\end{align}

and for $u=0$ we obtain,
\begin{align}
	\begin{split}
		&\eta^{mw}_{lt,u=0} =\frac{2-\beta_h\hbar\omega_c}{2+\beta_h\hbar\omega_c}
	\end{split}
\end{align}

The same as before, the efficiency at maximum power is maximum when $u\rightarrow 0$ and decreases with the velocity. Additionally, the quantum nature of the QHE is evident due to the fact that the $\hbar$ appears only through the combination $\hbar\omega$.

\subsection{Harmonic oscillator}

When it comes to the Harmonic oscillator as the working medium, the quantum Otto cycle has two isentropic and two isochoric strokes \cite{Kieu}, \cite{Quantum Afterburner}-\cite{Quan}. Following the same procedure as previously explained, the mean energies for every step are from \ref{energyHO}:
\begin{align}
	\begin{split}
		\langle \hat{H} \rangle_A &=\hbar\omega_c\Big(n(\omega_c,\beta_c)+\frac{1}{2}\Big), \langle \hat{H} \rangle_B =\hbar\omega_h\Big(n(\omega_c,\beta_c)+\frac{1}{2}\Big)\\
	\langle \hat{H} \rangle_C &=\hbar\omega_h\Big(N(\omega_h,\beta_h,u)+\frac{1}{2}\Big), \langle \hat{H} \rangle_D =\hbar\omega_c\Big(N(\omega_h,\beta_h,u)+\frac{1}{2}\Big)
	\end{split}
\end{align}

Moreover, the expectation value of heat and work are:

\begin{align}
\begin{split}
\langle \hat{W} \rangle_{AB} &= \hbar(\omega_h-\omega_c)\Big(n(\omega_c,\beta_c)+\frac{1}{2}\Big),\\
\label{Q2}
\langle \hat{Q} \rangle_{H}& =\hbar\omega_h\Big(N(\omega_h,\beta_h,u)-n(\omega_c,\beta_c)\Big),\\
\langle \hat{W} \rangle_{CD}&=-\hbar(\omega_h-\omega_c)\Big(N(\omega_h,\beta_h,u)+\frac{1}{2}\Big),\\
\langle \hat{Q} \rangle_{C}& = \hbar\omega_c\Big(n(\omega_c,\beta_c)-N(\omega_h,\beta_h,u)\Big).
\end{split}
\end{align}

\begin{align}
	\label{eta}
	\eta = -\frac{ \langle \hat{W} \rangle_{total}}{\langle \hat{Q} \rangle_{H}}=1-\frac{\omega_c}{\omega_h}.
\end{align}

We notice that the efficiency does not depend on the velocity of the moving hot bath.\\

\subsubsection{Total work}

Let us examine the {\em high temperature limit: $\beta \omega \rightarrow 0$}. The total produced work by the QHE is:

\begin{align}
	\begin{split}
		\langle W_{ht}\rangle&=\frac{1}{\beta_c}\Bigg(\frac{\eta}{1-\eta}-\frac{\sqrt{1-u^2}\log(\frac{1+u}{1-u})}{2u}\frac{\beta_c}{\beta_h}\eta\Bigg)\\
	\end{split}
\end{align}

when $u=0$ the result of Ref. \cite{Abah} is reproduced.

\begin{align}
	\begin{split}
		\langle W_{ht,u=0}\rangle=\frac{1}{\beta_c}\Bigg(\frac{\eta}{1-\eta}-\frac{\beta_c}{\beta_h}\eta\Bigg)\\
	\end{split}
\end{align}

\begin{figure}[H]
	\centering
	{{\includegraphics[width=15cm]{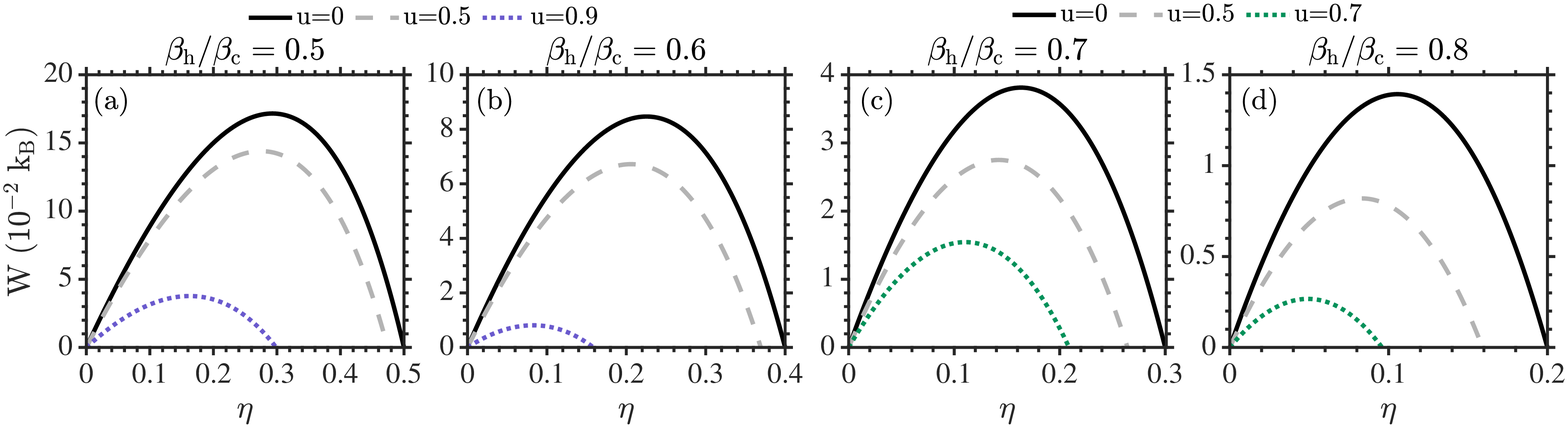} }}
	\caption{\small Total produced work with respect to efficiency for different ratios of temperatures of reservoirs $\frac{\beta_h}{\beta_c}$ and velocities of the heat bath.} 
\end{figure}

In Fig. 3, it is shown that the total produced work decreases with the velocity of the heat bath  and when $T_c\rightarrow T_h$.

Furthermore, the efficiency at maximum work, from (\ref{eta}) is

\begin{align}
	\begin{split}
		&\eta^{mw}_{ht} = 1-\frac{1}{(1-u^2)^{1/4}}\sqrt{\frac{2u}{\log(\frac{1+u}{1-u})}}\sqrt{\frac{\beta_h}{\beta_c}}
	\end{split}
\end{align}
for \begin{align}
	\omega_c=\omega_h \frac{1}{(1-u^2)^{1/4}}\sqrt{\frac{2u}{\log(\frac{1+u}{1-u})}}\sqrt{\frac{\beta_h}{\beta_c}}
\end{align}

and when $u=0$
\begin{align}
	\begin{split}
		&\eta^{mw}_{u=0} = 1-\sqrt{\frac{\beta_h}{\beta_c}}
	\end{split}
\end{align}
the Curzon-Ahlborn efficiency for classical systems \cite{Yair Rezek and Ronnie Kosloff}, \cite{Curzon}, \cite{Bihong Lin and Jincan Chen}, \cite{Deffner}  is reproduced.
\begin{figure}[H]
	\centering
	{{\includegraphics[width=7cm]{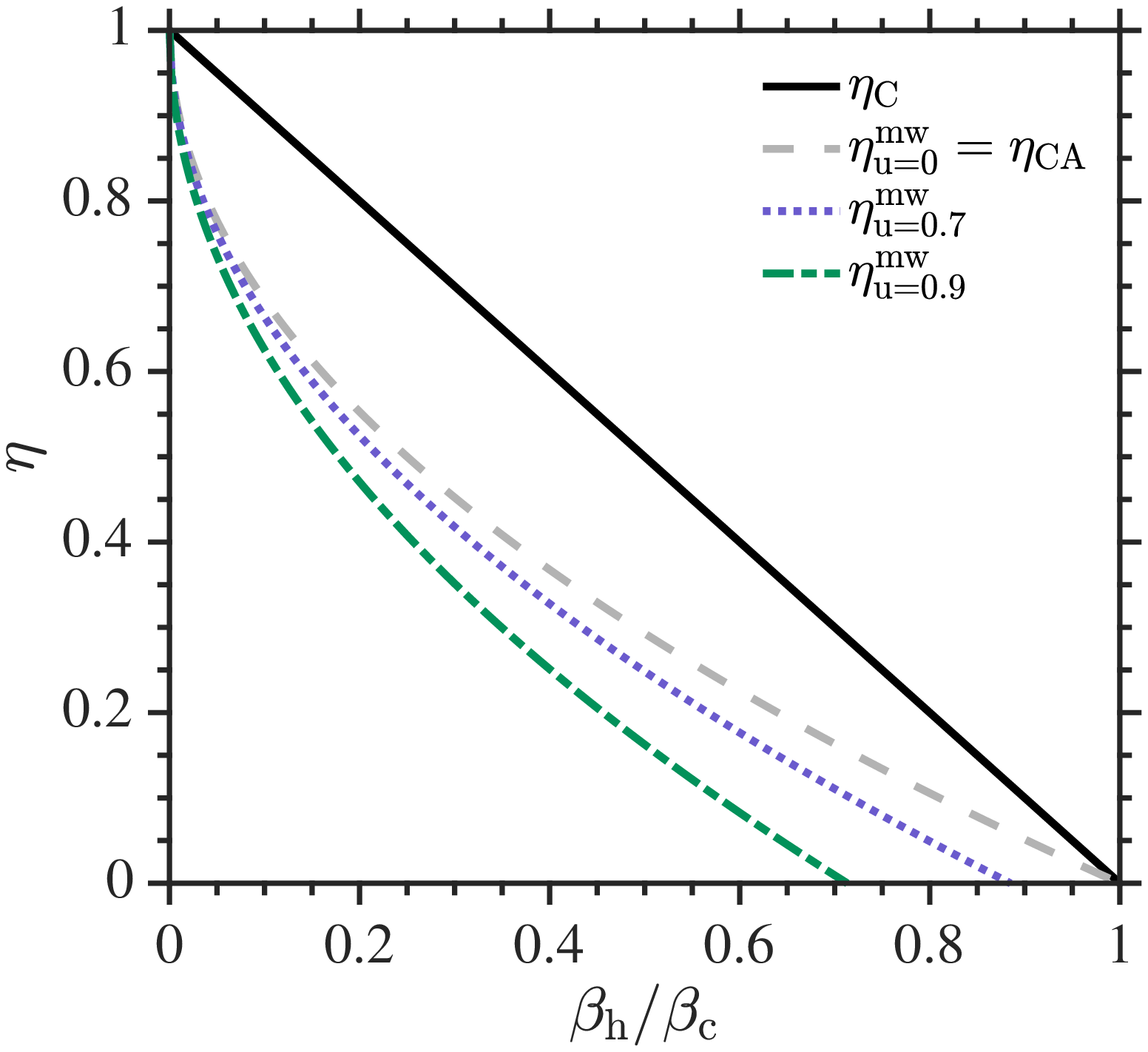} }}%
	% \caption{2 Figures}%
	\caption{ \small  Efficiency at maximum power with respect to $\beta_h/\beta_c$ in comparison with the Carnot limit and the efficiency of CA.}
\end{figure}

As indicated in Fig. 4 the efficiency at maximum power decreases with the velocity $u$.

Let us now, examine the {\em   low temperature limit: $\beta_c \omega_c \rightarrow \infty$}, the total produced work is:

\begin{align}
	\langle W_{lt}\rangle=\frac{(\omega_h-\omega_c)}{2}-\frac{\sqrt{1-u^2}\log(\frac{1+u}{1-u})}{2u\hbar}\frac{1}{\beta_h}\Big(1-\frac{\omega_c}{\omega_h}\Big)\\
\end{align}

The efficiency at maximum work, from (\ref{eta}) is

\begin{align}
	\begin{split}
		\label{efficiency low temp u}
		&\eta^{mw}_{lt} = 1-\frac{\sqrt{u}\sqrt{\beta_h}\sqrt{\hbar\omega_c}}{(1-u^2)^{1/4}\sqrt{\log(\frac{1+u}{1-u})} }
	\end{split}
\end{align}
for \begin{align}
	\begin{split}
		\omega_h=\sqrt{\omega_c} \frac{(1-u^2)^{1/4}\sqrt{\log(\frac{1+u}{1-u})} }{\sqrt{\hbar}\sqrt{u}\sqrt{\beta_h}}
	\end{split}
\end{align}
Moreover, when $u=0$ the result from \cite{Abah} is reproduced. 
\begin{align}
	\label{efficiency low temp}
	\begin{split}
		&\eta^{mw}_{lt,u=0} = 1-\sqrt{\frac{\beta_h\hbar\omega_c}{2}}
	\end{split}
\end{align}
In the low temperature regime the quantum character of  \ref{efficiency low temp u} and \ref{efficiency low temp} is evident because the $k_BT_c$ has been replaced from the ground state energy  $\frac{\hbar\omega_c}{2}$ of the quantum harmonic oscillator. In addition, the efficiency at maximum power decreases with the velocity.

\section{Effective Temperature}
\label{Effective Temperature}
In this section, we approach the issue of relativistic transformation of temperature  in the context of QHEs using a moving heat bath with relativistic velocity $u$ with respect to the cold bath. Thus, the aim is to define an effective temperature so that the efficiency at maximum power of QHE that uses a moving heat bath with relativistic velocity $u$, is equivalent to the efficiency at maximum power of QHE that uses a heat bath at rest. From previous results \cite{nikos}, we have shown that there is no relativistic rule for transformation of temperature, because a moving heat bath is
physically equivalent to a mixture of heat baths at rest, each with a different temperature.
 Consequently, we do not expect the definition of an effective temperature that could express the relativistic transformation of temperature. 

\section{Conclusions}
\label{Conclusions}
We analysed the quantum Otto thermodynamic cycle, using a moving heat bath with relativistic velocity $u$ with respect to the cold bath. This is accomplished by taking a small quantum system that is weakly coupled to a hot and a cold bath on alternate pace according to the Otto thermodynamic cycle. By using quasi-static reversible processes and the theory of open quantum system we define the condition for a close thermodynamic cycle. 

In both high and low temperature limits, we compute the total produced work, the efficiency and the efficiency at maximum power. The maximum efficiency of QHE does not depend on the velocity of the thermal reservoir. Moreover, using a classical thermal bath the same efficiency is found, which implies a universal bound that does not depend on the motion of the thermal reservoir.
 
  The total produced work and the efficiency at maximum power are affected negatively from the motion of the heat reservoir due to the fact that they decrease with the velocity of the thermal bath. This means that the motion of the heat bath has an energy cost for the QHE. Additionally, we notice that the efficiency at maximum power depends on the nature of the working medium. In particular, the QHE of two-level system surpasses the efficiency of CA. On the contrary, using as a working medium a harmonic oscillator, the efficiency at maximum power coincides with the CA efficiency. Moreover, when $T_c\rightarrow T_h$ the total produced work decreases. Moreover, the temperature determines the classical or quantum nature of the QHE. In particular, in the low temperature regime, the system has quantum nature because the thermodynamic quantities depend on $\hbar$ only through the combination $\hbar\omega$.
  
 Finally, It is impossible to define an effective temperature so that the efficiency at maximum power of QHE that uses a moving heat bath with relativistic velocity $u$, is equivalent to the efficiency at maximum power of QHE that uses a heat bath at rest.

In future work, we will extent this analysis to other thermodynamic cycles such as Carnot, Stirling, etc.

\section{ACKNOWLEDGMENTS}
I would like to thank my supervisor Charis Anastopoulos for his stimulating and fruitful discussions and suggestions that help me to complete this work. I would also like to thank Georgios Katsoulis for his valuable comments.


\begin{thebibliography}{9}
		


\bibitem{Alicki}
R. Alicki, {\em   The quantum open system as a model of the heat engine}J. Phys. A: Math. Gen. 12 (1979).


\bibitem{Abah0}
O. Abah and E. Lutz, {\em   When is a quantum heat engine Energy efficient quantum machines?}  EPL (Europhysics Letters) 118, 40005 (2017).


\bibitem{KosloffRezek}
R. Kosloff, and Y. Rezek, {\em   The quantum harmonic Otto cycle} Entropy 19, 136 (2017).


\bibitem{CHAND}
S. Chand and A. Biswas, {\em   Single-ion quantum Otto engine with always-on bath interaction}  EPL (Europhysics Letters) 118, 60003 (2017).



\bibitem{Abah2}
O. Abah and E. Lutz, {\em   Optimal performance of a quantum Otto refrigerator}  EPL (Europhysics Letters) 113, 60002 (2016).


\bibitem{Kosloff1}
R. Uzdin and R. Kosloff, {\em  Universal features in the efficiency at maximal work of hot quantum otto engines.}  EPL (Europhysics Letters) 118, 40001 (2014).


 

\bibitem{Beretta}
G. P. Beretta, {\em  Quantum thermodynamic Carnot and Otto-like cycles for a two-level system}  EPL (Europhysics Letters) 99, 20005 (2012).



\bibitem{Solfanelli}
A. Solfanelli, M. Falsetti, and M. Campisi, {\em Nonadiabatic single-qubit quantum Otto engine} Phys. Rev. B 101, 054513 (2020).





\bibitem{Mehta}
V. Mehta and R. S. Johal, {\em Quantum Otto engine with exchange coupling in the presence of level degeneracy} Phys. Rev. E 96, 032110 (2017).



\bibitem{Deffner}
S. Deffner, {\em Efficiency of harmonic quantum Otto engines at maximal power} Entropy 20, 875 (2018).




\bibitem{Leggio}
B. Leggio and M. Antezza, {\em Otto engine beyond its standard quantum limit} Phys. Rev. E 93, 022122 (2016).




\bibitem{reviews} C. K. Yuen, {\em  Lorentz Transformation of Thermodynamic Quantities}, Am. J. Phys. 38, 246 (1970); H. Callen and G. Horwitz, {\em Relativistic Thermodynamics}, Am. J. Phys. 39, 938 (1971); C. Farias, V. A. Pinto, and P. S.Moya, {\em What Is the Temperature of a Moving Body}, Sci. Rep. 7, 17657 (2017).
		

	
	
\bibitem{vMos} K. von Mosengeil, {\em Theorie der Stationären Strahlung in einem Gleichförmig Bewegten Hohlraum}, Ann. Phys. (Leipzig) 327,  867 (1907).
	
	
\bibitem{Planck} M. Planck, {\em Zur Dynamik Bewegter Systeme}, Sitzungsberichte der Königlich-Preussischen Akademie der Wissenschaften  542 (1907);
{\em  Zur Dynamik Bewegter Systeme}, Ann. Phys. 331, 1 (1908).
	
\bibitem{Einstein}A.  Einstein, {\em Über das Relativitätsprinzip und die aus Demselben Gezogenen Folgerungen},  Jahrb. Radioakt. Elektron. 4, 411 (1907).
	
\bibitem{Ott} H. Ott, {\em Lorentz-Transformation der W\"arme und der Temperatur}, Z. Physik 175, 70 (1963).
	
\bibitem{Arzel} H. Arzeliès, {\em Sur le Concept de Temperature en Thermodynamique Relativiste et en Thermodynamique Statistique}, Nuovo Ciment. B 40, 333 (1965).
	
\bibitem{Landsberg} P. T. Landsberg,  {\em Does a Moving Body Appear Cool?}, Nature 212, 571 (1966).
	
	
\bibitem{CaSa} G. Cavalleri and G.  Salgarelli, {\em Revision of the Relativistic Dynamics with Variable Rest Mass and Application to Relativistic Thermodynamics}, Nuovo Ciment. A62, 722 (1969).
	
 
			
\bibitem{Scovil}
H. E. D. Scovil and E. O. Schulz-DuBois, {\em Three-Level Masers as Heat Engines},  Phys. Rev. Lett. 2, 262 (1959).


	\bibitem{Klaers}
J. Klaers, S. Faelt, A. Imamoglu, and E. Togan, {\em Squeezed thermal reservoirs as a resource for a nano-mechanical engine beyond the Carnot limit}, Phys. Rev. X 7, 031044 (2017).	

 

\bibitem{ZhangGuo}
Y. Zhang, J. Guo and J. Chen, {\em Unified trade-off optimization of quantum Otto heat engines with squeezed thermal reservoirs}, Quantum Inf Process 19, 268 (2020).




\bibitem{Zhang}
Y. Zhang, {\em Optimization performance of quantum Otto heat engines and refrigerators with squeezed thermal reservoirs}, Physica A, 559, 125083 (2020).



	\bibitem{Manzano}
G. Manzano, F. Galve, R. Zambrini, and J. M. R. Parrondo, {\em 	Entropy production and thermodynamic power of the squeezed thermal reservoir}, Phys. Rev. E 93, 052120 (2016).
	\bibitem{Assis}
R. J. Assis, J. S. Sales, J. A. R. Cunha, and N. G. Almeida, {\em 	Universal two-level quantum Otto machine under a squeezed reservoir}, Phys. Rev. E 102, 052131 (2020).


\bibitem{Abah1} 
O. Abah and E. Lutz, {\em Efficiency of heat engines coupled to nonequilibrium reservoirs}, EPL (Europhysics Letters) 106, 20001 (2014).

	\bibitem{Alickiengine} 
D. Gelbwaser-Klimovsky, R. Alicki, and G. Kurizki, {\em Minimal universal quantum heat machine}, Phys. Rev. E 87, 012140  (2013).

	\bibitem{Scullycoherent} 
M. O. Scully, M. S. Zubairy, G. S. Agarwal, and H.Walther, {\em Extracting work from a single heat bath via vanishing quantum coherence}, Science 299, 862 (2003).



\bibitem{Dillenschneidercorrelation} 
R. Dillenschneider and E. Lutz, {\em  Energetics of quantum correlations},  Europhys. Lett.) 88, 50003 (2009).


\bibitem{nikos} 
N. Papadatos and C. Anastopoulos, {\em Relativistic Quantum Thermodynamics of Moving Systems}, Phys. Rev. D 102, 085005  (2020).
	
	
	
\bibitem{BrePe07} 
H. P. Breuer and F. P. Petruccione, {\em The Theory of Open Quantum Systems} (Oxford University Press, 2007).
	
	
\bibitem{Eitan Geva and Ronnie Kosloff}
E. Geva and R. Kosloff, {\em   A quantum-mechanical heat engine operating in finite time. A model consisting of spin-1/2 systems as the working fluid} J. Chem. Phys. 96, 3054 (1992)
	
	
	
	
\bibitem{Yair Rezek and Ronnie Kosloff}
Y. Rezek and R. Kosloff, {\em Irreversible performance of a quantum harmonic heat engine}, New J. Phys. 8, 83 (2006).
	
	
	
\bibitem{entropyproductionLee}
J. S. Lee, S. H. Lee, Jaegon Um and Hyunggyu Park, {\em Carnot efficiency and zero-entropy-production rate do not guarantee reversibility of a process}, Journal of the Korean Physical Society volume 75, (2019).
	
	
	
\bibitem{Callen}
H. B. Callen, {\em Thermodynamics and an Introduction to Thermostatistics}, Wiley: New York, NY, USA, (1985).
	
	\bibitem{Curzon}
F. L. Curzon and B. Ahlborn, {\em  Efficiency of a carnot engine at maximum power output}, Am. J. Phys. 43, 22 (1975).
	
	
\bibitem{Broeck}
C. Van den Broeck, {\em  Thermodynamic Efficiency at Maximum Power}, Phys. Rev. Lett. 95, 190602 (2005).

	
	\bibitem{Kieu}
T. D. Kieu,  {\em  	The Second Law, Maxwell’s Demon, andWork Derivable from Quantum Heat Engines}, Phys. Rev. Lett. 93, 140403 (2004).



\bibitem{Abah}
O. Abah, J. Roßnagel, G. Jacob, S. Deffner, F. Schmidt-Kaler, K. Singer, and E. Lutz {\em  Single-Ion Heat Engine at Maximum Power}, Phys. Rev. Lett. 109, 203006 (2012).


\bibitem{Wel00} H. A. Weldon, {\em Thermal Green Functions in Coordinate Space for Massless Particles of any Spin}, Phys.Rev. D62,  056010 (2000).

	
\bibitem{Letaw}J. R. Letaw, {\em Stationary World Lines and the Vacuum Excitation of Noninertial Detectors}, Phys. Rev. D 23, 1709 (1981). 

	
\bibitem{Unruh76} W. G. Unruh, {\em Notes on Black Hole Evaporation}, Phys. Rev. D14, 870 (1976).
	
	
\bibitem{Dewitt}B. S. DeWitt, {\em Quantum Gravity: the New Synthesis} in
General Relativity: An Einstein Centenary Survey, ed. by S. W. Hawking
and W. Israel (Cambridge University Press, Cambridge 1979), p.680.
	
\bibitem{HLL12} B. L. Hu, S-Y Lin, J. Louko, {\em Relativistic Quantum Information in Detectors–Field Interactions}, Class. Quantum Grav. 29, 224005 (2012).
	
\bibitem{CoMa} S. S. Costa and G. E. A. Matsas, {\em Temperature and Relativity}, Phys. Lett. A209, 155 (1995).
		
\bibitem{LaMa}  P. T. Landsberg  and G. E. A. Matsas, {\em Laying the Ghost of the Relativistic Temperature Transformation}, Phys. Lett. A223, 401 (1996).
	
\bibitem{LaMa1}  P. T. Landsberg  and G. E. A. Matsas, {\em The impossibility of a universal relativistic temperature transformation}, Physica A 340, 92 (2004).
	
	
\bibitem{Naka}
T. K. Nakamura, {\em Lorentz Transform of Black-Body Radiation Temperature},  Europhys. Lett. 88, 20004  (2009).
	
	
	
\bibitem{Andresen} 
B. Andresen, P. Salamon, and R. S. Berry, {\em Thermodynamics in finite time} Phys. Today 37, No. 9, 62 (1984).
		
\bibitem{Quantum Afterburner}
M. O. Scully, {\em Quantum Afterburner: Improving the Efficiency of an Ideal Heat Engine}, Phys. Rev. Lett. 88, 050602 (2002).
	
	
\bibitem{Bihong Lin and Jincan Chen}
B. Lin and J. Chen, {\em Performance analysis of an irreversible quantum heat engine working with harmonic oscillators}, Phys. Rev. E 67, 046105 (2003).
	
	
	
\bibitem{Quan}
H. T. Quan, Yu-xi Liu, C. P. Sun, and Franco Nori, {\em Quantum thermodynamic cycles and quantum heat engines}, Phys. Rev. E 76, 031105 (2007).
	
	
	
\bibitem{Bihong Lin and Jincan Chen}
B. Lin and J. Chen {\em Performance analysis of an irreversible quantum heat engine working with harmonic oscillators}, Phys. Rev. E 67, 046105 (2003).
	
	
	
	
\bibitem{Deffner}
S. Deffner, {\em  Efficiency of Harmonic Quantum Otto Engines at Maximal Power}, Entropy, 20, 875 ( 2018).
	

\end{thebibliography}
\end{document}